\documentclass[twocolumn,aps,showpacs,prb,tightenlines,amsmath,amssymb,superscriptaddress]{revtex4}
\usepackage{graphicx}
\usepackage{amssymb}
\usepackage{dcolumn}
\usepackage{amsmath}
\usepackage{bm}% bold math

\usepackage{colordvi}
\newcommand{\bgreek}[1]{\mbox{\boldmath$#1$\unboldmath}}

\begin{document}

\title{Existence of vertical spin stiffness in
  Landau-Lifshitz-Gilbert equation in ferromagnetic semiconductors}

\author{K.\ Shen}
\affiliation{Hefei National Laboratory for Physical Sciences at
  Microscale and Department of Physics, 
University of Science and Technology of China, Hefei,
  Anhui, 230026, China}
\author{G.\ Tatara}
\affiliation{Department of Physics, Tokyo Metropolitan University,
Hachioji, Tokyo 192-0397, Japan}
\author{M.\ W.\ Wu}
\thanks{Author to whom correspondence should be addressed}
\email{mwwu@ustc.edu.cn.}
\affiliation{Hefei National Laboratory for Physical Sciences at
  Microscale and Department of Physics, 
University of Science and Technology of China, Hefei,
  Anhui, 230026, China}
\date{\today}

\begin{abstract}
We calculate the magnetization torque due to the spin polarization of
the itinerant electrons by deriving the kinetic spin Bloch equations
based on the $s$-$d$ model. We find that the first-order gradient of the
magnetization inhomogeneity gives rise to the current-induced torques, which are
consistent to the previous works. At the
second-order gradient, we find an effective magnetic field
perpendicular to the spin stiffness filed. This field is proportional to
the nonadiabatic parameter $\beta$. We show that this
vertical spin stiffness term can significantly modify the domain-wall
structure in ferromagnetic semiconductors and hence should be included in
the Landau-Lifshitz-Gilbert equation in studying the
magnetization dynamics. 
\end{abstract}

%%75.60.Ch 	Domain walls and domain structure (for magnetic
%%              bubbles and vortices, see 75.70.Kw)
%%72.25.Dc 	Spin polarized transport in semiconductors 
%%75.30.Gw 	Magnetic anisotropy 
%%75.50.Pp 	Magnetic semiconductors
\maketitle
\section{Introduction}
Ferromagnetic systems have attracted much interest for a long history
because of the intriguing physics and applications.\cite{kittel,hubert}
As the development of information technology, the research on
magnetization dynamics in micromagnets has become an active
field.\cite{tserk,tatara} Great efforts have been devoted to this field by
aiming to manipulate magnetization more efficiently.\cite{allwood,thomas,parkin} For theoretical
simulation, the magnetization dynamics is usually described by the
Landau-Lifshitz-Gilbert (LLG) equation,\cite{gilbert,landau}
\begin{equation}
  {\dot {\bf n}}=-\gamma{\bf n}\times{\bf H}_{\rm eff}+\alpha{\bf
    n}\times{\dot {\bf n}}-(1-\beta {\bf n}\times)({\bf
    v}_s\cdot\nabla){\bf n},
  \label{eq1}
\end{equation}
where ${\bf n}$ represents the direction of the magnetization. ${\bf
  H}_{\rm eff}$ in the first term on the right-hand side of
Eq.\,(\ref{eq1}) is the effective magnetic field which drives the magnetization
procession and determines the domain structure in the equilibrium
states. Up to date, different sources of effective magnetic
field have been identified, e.g., the external 
magnetic field, the crystal anisotropy induced by the spin-orbit
coupling, the exchange energy due to the 
spatial inhomogeneity, and also the demagnetization field from
dipole-dipole interaction.\cite{tserk2}
The second term, the Gilbert damping torque, describes the
magnetization relaxation to the effective field axis on the time scale of
$1/(\alpha\gamma H_{\rm eff})$.\cite{gilbert}
The last one with first-order gradient of the magnetization is the spin
torque induced by the transport of the 
spin polarized itinerant electrons, where
${\bf v}_s$ is proportional to the spin current
density.\cite{berger,szhang,tserk4,thia,kohno1,kohno2,szhang2} The 
$\beta$-term,\cite{szhang,thia} first proposed by Zhang and Li,\cite{szhang} was
demonstrated to be critically important to the current-driven domain
wall motion, which overwhelms the threshold current due to
pinning force and transverse anisotropy for domain wall
motion\cite{tatara2,tatara3,tserk2} and results in the steady domain wall velocity $\propto
\beta/\alpha$ in the absence of the external magnetic
field. Therefore, the determination of the effective magnetic field
and the parameters in the LLG
equation, such as $\alpha$ and $\beta$, is an important issue for
magnetization dynamics study.\cite{szhang,kohno1,kohno2,hals,shen}

Previously, we have derived the Gilbert damping ($\alpha$) term based
on the kinetic spin Bloch equation (KSBE) approach\cite{wu} in homogeneous ferromagnetic
systems.\cite{shen} In the present work, we derive the whole LLG equation 
from the $s$-$d$ model in inhomogeneous ferromagnetic systems based on
the same approach.
From the first-order gradient of the magnetization inhomogeneity, we
obtain the current-induced torques which are consistent to the above
LLG equation. Within the second-order gradient, we find that the LLG
equation should be modified and written as
\begin{eqnarray}
  \nonumber
  \dot{\bf n}
  &=&-\gamma {\bf n}\times {\bf  H}_{\rm
    eff}+{\alpha}{\bf n}\times{\dot{\bf n}}
-(1-{\beta}{\bf n}\times)({\bf v}_s\cdot
  \nabla){\bf n}\\
  &&\hspace{-0.1cm}\mbox{}
  -\tfrac{\gamma}{M_d} A_{\rm ss}{\bf n}\times(1-{\beta} {\bf
    n}\times)\nabla^2{\bf n}.
  \label{eq2}
\end{eqnarray}
The second-order gradient introduces two contributions to the
effective magnetic field.
The one given by $A_{\rm ss}\nabla^2{\bf n}$ is
identified as the spin stiffness field discussed in previous
works.\cite{tserk2,konig,konig2}  The
other term, in the form of $-\beta A_{\rm ss}{\bf n}\times
\nabla^2{\bf n}$, has never be referred in the literature. In the
present paper, we call it ``vertical spin stiffness'' in the sense
of the fact that this new field is vertical to the plane defined by
the magnetization and the normal spin stiffness $\nabla^2{\bf n}$. Interestingly, this
vertical spin stiffness can not be written in terms of
the free energy, and therefore, it can not be derived from
the functional derivative of the free energy 
with respect to the local magnetization ${\bf H}_{\rm eff}=-\delta
F[{\bf M}_d]/{{\bf M}_d}$ previously.\cite{landau,tserk2} We
find that this vertical spin stiffness results in the tilt of the
magnetization. The new term can significantly change the domain-wall
structure in ferromagnetic semiconductors. Since the magnitude of this
field is proportional to the factor of $\beta$, the proposed effect is expected to be
important in ferromagnetic semiconductors where $\beta$ is large
due to the strong spin-orbit interaction.\cite{hals,jawo}

This paper is organized as follows: In Sec.\,II, we setup our model and
derive the KSBEs for the itinerant
electrons in the inhomogeneous ferromagnetic system. We calculate
the spin torque by solving the KSBEs in Sec.\,III and discuss the
results in Sec.\,IV. Finally, we briefly summarize in Sec.\,V.
\section{KSBEs}
 We use the exchange interaction  
Hamiltonian density $H_{\rm sd}({\bf r})=M{\bf
  n}\cdot {\bgreek \sigma}$ with $M$ denoting the coupling
constant. Following Ref.\,{\onlinecite{nagaev}}, we here assume that the
ferromagnetic interaction exists among ${\bf n}({\bf r})$ and show later that this
assumption is justified after integrating out the conduction electrons
(giving rise to spin stiffness).
The Pauli matrices ${\bgreek \sigma}$ are used to describe the
itinerant electrons. In contrast to the previous work on the Gilbert
damping,\cite{shen} we introduce the inhomogeneity by considering
the position dependence of the magnetization direction ${\bf n}({\bf
  r})={\bf M}_d({\bf r})/M_d$. $M_d$ is the uniform saturate
magnetization. For the strong exchange coupling in ferromagnetic
system, the rotation framework\cite{tatara,kohno2} is employed here. That is, the
local spinor operators of the itinerant electrons are defined as
$a({\bf r})=(a_\uparrow({\bf r}),a_\downarrow({\bf r}))^T$, with
$\uparrow$ ($\downarrow$) labeling the spin orientation parallel
(anti-parallel) to ${\bf n}({\bf r})$. Therefore, one has ${H}_{\rm
  sd}=Ma^\dag\sigma_za$. The spinor operators $a({\bf r})$ are connected to
the ones defined in the lattice coordinate system
$c=(c_\uparrow,c_\downarrow)^T$ via 
the unitary transformation $a({\bf r})=U({\bf r})c$. The
transformation matrices is given by $U({\bf r})={\bf
  m}({\bf r})\cdot {\bgreek\sigma}$ with ${\bf m}=({\rm sin}(\theta/2){\rm
cos}\varphi,{\rm sin}(\theta/2){\rm sin}\varphi,{\rm
cos}\theta)$ for ${\bf n}({\bf r})=
({\rm sin}\theta{\rm cos}\varphi,{\rm sin}\theta{\rm sin}\varphi,{\rm
  cos}\theta)$.\cite{tatara,kohno2} 

In the rotation framework, the kinetic Hamiltonian should be expressed as
${ H}_{\rm k}=|\nabla c|^2/2m=|(\nabla+i{\bf A})a|^2/2m=|\nabla a|^2/2m+H_A$, where
the gauge field introduced by the coordinate transformation is given
by ${A}_i=-iU^\dag\partial_i U=({\bf
  m}\times\partial_i{\bf m})_l\sigma_l=A^l_i\sigma_l$. 
Then, one obtains the Hamiltonian density associated with
the gauge field,\cite{tatara}
\begin{equation}
   { H}_A=-\frac{i}{2m}\sum_i[a^\dag A_i \nabla_i a-(\nabla_i
  a)^\dag A_i a]+a^\dag \frac{A^2}{2m} a.
   \label{eq3}
\end{equation}

To derive the KSBEs of the itinerant
electrons, we follow the nonequilibrium Green function
approach.\cite{wu,wu2,haug} The Dyson equation of the contour-ordered
Green function can be expressed as
\begin{eqnarray}
  \nonumber
  G(1,2)&=&G_0(1,2)+\int_c d3G_0(1,3)U_e(3)G(3,2)\\
  \nonumber
  &&\hspace{-0.1cm}\mbox{}+\int_c d3\int_c d4G_0(1,3)\Sigma(3,4)G(4,2)\\
  \nonumber
  &=&G_0(1,2)+\int_c d3G(1,3)U(3)G_0(3,2)\\ 
  &&\hspace{-0.1cm}\mbox{}+\int_c d3\int_c
  d4G(1,3)\Sigma(3,4)G_0(4,2),
  \label{eq4}
\end{eqnarray}
where the single particle contour-ordered Green function between two
space-time points $(1)=({\bf r_1},t_1)$ and $(2)=({\bf r}_2,t_2)$ on
the contour $C$ is defined as
$G(1,2)=-i\langle T_C[\psi_H(1)\psi_H^\dag(2)]\rangle$.\cite{haug}
$U_e$ describes the local electric potential energy,
whereas $\Sigma$ represents the self-energy correction due to
interactions, such as the electron-impurity, electron-phonon, and
electron-electron interactions. $G_0$ stands for the free-particle Green function.
The left- and right-inverses of $G_0(1,2)$ are given by
\begin{eqnarray}
 (G_0(x_1,x_2))_{x_1}^{-1}&=&i\partial_{t_1}-H_0({\bf p}_1,{\bf r}_1),\\
 (\stackrel{\leftarrow}{G}_0(x_1,x_2))_{x_2}^{-1}&=&-i\stackrel{\leftarrow}{\partial}_{t_2}
 -H_0(-\stackrel{\leftarrow}{\bf p}_2,{\bf r}_2),
 \label{eq5-6}
\end{eqnarray}
with $H_0=H_{\rm k}+H_{\rm sd}$. By multiplying them to
Eq.\,(\ref{eq4}), one obtains
\begin{eqnarray}
  \nonumber
  &&\hspace{-0.5cm}i(\partial_{t_1}+\partial_{t_2})G(1,2)=[H_0({\bf p}_1,{\bf r}_1)+U_e(1)]G(1,2)\\
  \nonumber
  &&\mbox{}-G(1,2)[H_0(-\stackrel{\leftarrow}{\bf p}_2,{\bf r}_2)+U_e(2)]\\
  &&\mbox{}+\int_c d3 [\Sigma(1,3)G(3,2)-G(1,3)\Sigma(3,2)],
  \label{eq7}
\end{eqnarray}
with ${\bf p}_i$ corresponding to the momentum
operators $-i\partial_{{\bf r}_i}$. 
To transform the above equation to the center-of-mass and
relative variables
\begin{eqnarray}
  \nonumber
  & {\bf R}=({\bf r}_1+{\bf r}_2)/2; & {\bf r}={\bf r}_1-{\bf r}_2;\\
  & T=(t_1+t_2)/2; & t=t_1-t_2,
  \label{eq8}
\end{eqnarray}
one rewrites the following Green functions as
\begin{eqnarray}
  G(1,2)&=&G({\bf R},{\bf r},T+\tfrac{t}{2},T-\tfrac{t}{2}),\\
  G(1,3)&=&e^{\frac{{\bf r}_3-{\bf r}_2}{2}\partial_{\bf R}}G({\bf
    R},{\bf r}_1-{\bf r}_3,T+\tfrac{t}{2},t_3),\\
  G(3,2)&=&e^{\frac{{\bf r}_3-{\bf r}_1}{2}\partial_{\bf R}}G({\bf
    R},{\bf r}_3-{\bf r}_2,t_3,T-\tfrac{t}{2}).
  \label{eq2229-11}
\end{eqnarray}
The self-energy can be written in the same way.
Similarly, one obtains
\begin{eqnarray}
  U_e(1)&=&e^{\frac{\bf r}{2}\partial_{\bf R}}U_e({\bf
    R},T+\tfrac{t}{2}),\\
  U_e(2)&=&e^{-\frac{\bf r}{2}\partial_{\bf R}}U_e({\bf
    R},T-\tfrac{t}{2}).
  \label{eq22212-13}
\end{eqnarray}
The Hamiltonian can be written as
\begin{eqnarray}
H_0({\bf p}_1,{\bf r}_1)&=&e^{\frac{\bf r}{2}\partial_{\bf R}^{H_0}}H_0(\tfrac{1}{2}{\bf
    P}_{\bf R}+{\bf p},{\bf R}),\\
H_0(-\stackrel{\leftarrow}{\bf p}_2,{\bf r}_2)&=&H_0(-\tfrac{1}{2}\stackrel{\leftarrow}{\bf 
    P}_{\bf R}+\stackrel{\leftarrow}{\bf p},{\bf R})e^{-\frac{\bf
    r}{2}{\stackrel{\longleftarrow}{\partial_{\bf R}^{H_0}}}},
  \label{eq22214-15}
\end{eqnarray}
where ${\bf P}_{\bf R}$ and ${\bf p}$ represent the momentum operators
respect to the center-of-mass and relative variables.
According to Eq.\,(\ref{eq3}), the left- and right-operators of the
kinetic Hamiltonian should be different and can be written as
\begin{eqnarray}
  H_{\rm k}({\bf p},{\bf r})&=&\tfrac{1}{2m}[{\bf p}^2-i(\nabla \cdot
  {\bf A})+2{\bf A}\cdot {\bf p}+{\bf A}^2],\\
  {H}_{\rm k}(-\stackrel{\leftarrow}{\bf p},{\bf
    r})&=&\tfrac{1}{2m}[\stackrel{\leftarrow}{\bf p}^2+i(\nabla \cdot {\bf  
    A})-2\stackrel{\leftarrow}{\bf p}\cdot {\bf 
    A}+{\bf A}^2],
\label{eq22216-17}
\end{eqnarray}
separately. By assuming the magnetization varies smoothly respect to
the spatial coordinates, we only keep the spatial gradient up to the
second order. Since the gauge field is already the first-order gradient,
both $\nabla\cdot {\bf A}$ and ${\bf A}^2$ are second-order gradient
terms. We include both of them and neglect the higher-order ones.
The gradient term of the gauge field can be written in the center-of-mass 
coordinate system as
\begin{eqnarray}
  \nonumber
  \nabla_{{\bf r}_1}\cdot {\bf A}({\bf r}_1)
  &=&\tfrac{1}{2}\nabla_{\bf R}\cdot{\bf A}({\bf R}+\tfrac{\bf
    r}{2})+\nabla_{\bf r}\cdot {\bf  A}({\bf R}+\tfrac{\bf r}{2})\\
  \nonumber
  &=& \tfrac{1}{2}\nabla_{\bf R}\cdot{\bf
    A}({\bf R})+\partial_{r_i}\tfrac{r_j}{2}\partial_{R_j} A^i({\bf R})\\
  &=&\nabla_{\bf R}\cdot {\bf A}({\bf R}).
  \label{eq18}
\end{eqnarray}
Similar calculation gives
\begin{equation}
  \nabla_{{\bf r}_2}\cdot {\bf A}({\bf
    r}_2)=\nabla_{\bf R}\cdot {\bf A}({\bf R}).
  \label{eq19}
\end{equation}
Moreover, one can easily show that ${\bf A}^2({\bf r}_i)\approx{\bf A}^2({\bf
  R})={\bf I}\sum_{i,l}(A_i^l)^2$
with ${\bf I}$ representing the unit matrix.

By substituting all these equations into Eq.\,(\ref{eq7}) and doing
Fourier transformation respect to the relative coordinate ${\bf r}$,
one obtains
\begin{eqnarray}
  \nonumber
  &&\hspace{-0.0cm}i\partial_TG({\bf R},{\bf
    k},t_1,t_2)\\
  \nonumber
  &&=e^{i\frac{1}{2}(\partial_{\bf k}\partial_{\bf
      R}^{H_0}-\partial_{\bf R}^G\partial_{\bf k}^{H_0})}H_0({\bf
    k},{\bf R})G({\bf R},{\bf k}, t_1,t_2)\\
  \nonumber
  &&\hspace{0.2cm}\mbox{}-e^{-i\frac{1}{2}(\partial_{\bf k}\partial_{\bf
      R}^{H_0}-\partial_{\bf R}^G\partial_{\bf k}^{H_0})}G({\bf
    R},{\bf k},t_1,t_2) H_0({\bf k},{\bf R}) \\
  \nonumber
  &&\hspace{0.2cm}\mbox{}+e^{i\frac{1}{2}\partial^G_{\bf k}\partial_{\bf
      R}^{U_e}}U_e({\bf R},t_1)G({\bf R},{\bf k},t_1,t_2)\\
  \nonumber
  &&\hspace{0.2cm}\mbox{}-e^{-i\frac{1}{2}\partial^G_{\bf k}\partial_{\bf R}^{U_e}}G({\bf R},{\bf
    k},t_1,t_2)U_e({\bf R},t_2)\\
  \nonumber
  &&\hspace{0.2cm}\mbox{}+\int_c dt_3[e^{\tfrac{i}{2}(\partial^G_{\bf
      k}\partial^\Sigma_{\bf R}-\partial_{{\bf
        k}}^\Sigma\partial^G_{\bf R})}
  \Sigma({\bf R},{\bf k},t_1,t_3)G({\bf R},{\bf
    k},t_3,t_2)\\
  &&\hspace{0.2cm}\mbox{}-e^{\tfrac{i}{2}(\partial^\Sigma_{\bf
      k}\partial^G_{\bf R}-\partial_{{\bf
        k}}^G\partial^\Sigma_{\bf R})}
  G({\bf R},{\bf k},t_1,t_3)\Sigma({\bf R},{\bf
    k},t_3,t_2)].
  \label{eq20}
\end{eqnarray}
The details can be found in Appendix\,A.
We then perform the gradient expansion up to the first order and obtain
\begin{eqnarray}
  \nonumber
  i\partial_TG
  &=&[H_0+U_e, G]-\tfrac{i}{2}\{\partial_{\bf k}H_0, \partial_{\bf R} G\}
  +\tfrac{i}{2}\{\partial_{\bf R}\partial_{\bf k}H_0,G\}\\
  &&\hspace{-0.2cm}\mbox{}+\tfrac{i}{2}\{\partial_{\bf R}(H_0+U_e),\partial_{\bf k}G\}
  +\int_c dt_3(\Sigma G-G\Sigma),
  \label{eq21}
\end{eqnarray}
where all the quantities are defined at ${\bf R}$ and ${\bf k}$.
We should point out that the commutator notation $[H_0+U_e, G]$ is still
used although the left- and right-operators of $H_0$ are in different expressions
[see Eqs.\,(16) and (17)].
By taking the isochronous condition, i.e., $t\to 0$, one has\cite{haug}
\begin{eqnarray}
  \nonumber
  &&\int_c dt_3 [\Sigma(T,\tau) G(\tau,T)-G(T,\tau)\Sigma(\tau,T)]^<\\
  \nonumber
  &&=\int_{-\infty}^Td\tau
  [\Sigma^>(T,\tau)G^<(\tau,T)
  -\Sigma^<(T,\tau)G^>(\tau,T)\\
  &&\hspace{0.2cm}\mbox{}-G^>(T,\tau)\Sigma^<(\tau,T)
  +G^<(T,\tau)\Sigma^>(\tau,T)],
  \label{eq22}
\end{eqnarray}
where the lesser and greater Green functions are defined by
$G^<(\tau,\tau^\prime)=i\langle\psi_H^\dag(\tau^\prime)\psi_H(\tau)\rangle$ and
$G^>(\tau,\tau^\prime)=-i\langle\psi_H(\tau)\psi_H^\dag(\tau^\prime)\rangle$.
Therefore, the correlation function $G^<(T,T)$ 
can be written as 
\begin{eqnarray}
  \nonumber
  i\partial_TG^<
  &=&\int_{-\infty}^T d\tau(\Sigma^> G^<-\Sigma^<G^>-G^>\Sigma^<+G^<\Sigma^>)  \\
  \nonumber
  &&\hspace{-0.2cm}\mbox{}+\tfrac{i}{2}\{\partial_{\bf R}\partial_{\bf k}H_0,G^<\}
  +\tfrac{i}{2}\{\partial_{\bf R}(H_0+U_e),\partial_{\bf k}G^<\}\\
  &&\hspace{-0.2cm}\mbox{}+[H_0+U_e, G^<]-\tfrac{i}{2}\{\partial_{\bf k}H_0, \partial_{\bf R} G^<\}.
  \label{eq23}
\end{eqnarray}
Within the generalized Kadanoff-Baym ansatz,\cite{haug} we have $G^<({\bf R},{\bf
  k},T,T)=i\rho_{\bf k}({\bf R},T)$
where $\rho_{\bf k}({\bf R}, T)$ is the local density matrix of the itinerant
electrons with momentum $\bf k$ located at ${\bf R}$. Therefore, we
write the general form of the KSBEs of
the itinerant electrons as
\begin{eqnarray}
  \nonumber
  &&\partial_t\rho_{\bf k}+i[H_0, \rho_{\bf k}]+\tfrac{1}{2}\{
  \nabla_{\bf k} H_0, \nabla_{\bf R}\rho_{\bf  
    k}\}-\nabla_{\bf R}U_e\cdot \nabla_{\bf k}\rho_{\bf
    k}\\
  \nonumber
  &&\mbox{}-\tfrac{1}{2}\{\nabla_{\bf R}H_0,\nabla_{\bf k}\rho_{\bf
    k}\}-\tfrac{1}{2}\{\nabla_{\bf R}\cdot\nabla_{\bf k}H_0, \rho_{\bf k}\}\\
  &&=\partial_t\rho_{\bf k}|^c_{\rm scat}+\partial_t\rho_{\bf k}|^f_{\rm scat}.
\label{eq24}
\end{eqnarray}
On the right-hand side of above equations, $\partial_t\rho_{\bf k}|^c_{\rm
  scat}$ and $\partial_t\rho_{\bf k}|^f_{\rm
  scat}$ from the integral term in Eq.\,(\ref{eq23}) represent the
spin-conserving and spin-flip scatterings. The details of these terms can be
found in Ref.\,\onlinecite{wu}.

We specify the Hamiltonian
$H_0=M\sigma_z+[(k^2+2A_ik_i+A_i^lA_i^l){\bf I}\mp i(\nabla\cdot{\bf
  A})]/(2m)$ with the upper (lower) sign 
representing the left (right) operator case.
The electric potential energy is given by $U_e=e{\bf
  E}\cdot{\bf R}$ ($e>0$). The final form of the KSBEs is given by 
\begin{eqnarray}
  \nonumber
  &&\partial_t\rho_{\bf k}+i[M\sigma_z, \rho_{\bf
    k}]+i\tfrac{k_i}{m}[A_i,\rho_{\bf k}]+\tfrac{1}{2m}\{{\bf A},\nabla_{\bf R}\rho_{\bf k}\}\\
  \nonumber
  &&\mbox{}+\tfrac{1}{m}{\bf k}\cdot \nabla_{\bf R}\rho_{\bf 
    k}-e{\bf E}\cdot \nabla_{\bf k}\rho_{\bf k}
  -\tfrac{k_i}{2m}\{\nabla_{\bf k}\rho_{\bf k},\nabla_{\bf R}A_i\}\\
  &&=\partial_t\rho_{\bf k}|^c_{\rm scat}+\partial_t\rho_{\bf
    k}|^f_{\rm  scat}.
\label{eq25}
\end{eqnarray}
Interestingly, we find that the contribution from $-\tfrac{1}{2}\{\nabla_{\bf
  r}\cdot\nabla_{\bf k}H_0, \rho_{\bf k}\}$ is completely canceled by the
gauge field gradient term from $i[H_0, \rho_{\bf k}]$ and the ${\bf
  A}^2$-term is irrelevant.

\section{Solution of KSBEs}
In general cases, the KSBEs are too complicated to solve
analytically and the numerical scheme should be employed.
However, the analytical solution can still be expected within some
simplification of the KSBEs. In the following, we assume:
({\em i}) the spatial dependence is weak in the rotation coordinate systems,
hence $\nabla_{\bf R}\rho_{\bf k}\approx 0$; ({\em ii}) the scattering is
strong enough to set up the steady-state condition $\partial_t\rho_{\bf k}\approx
0$. Without loss of generality, we take the 
magnetization gradient along arbitrary direction.
The external electric field is applied on the
purpose of producing current-induced magnetization
dynamics. Therefore, the KSBEs become 
\begin{eqnarray}
 \nonumber
 &&-e{\bf E}\cdot\nabla_{\bf k}\rho_{\bf
   k}+i[\tfrac{k_i}{m}A_i+M\sigma_z,\rho_{\bf
   k}]-\tfrac{k_i}{2m}\{\nabla_{\bf k}
 \rho_{\bf k},\nabla_{\bf R}A_i\} \\
 &&=\partial_t\rho_{\bf k}|^c_{\rm scat}+\partial_t\rho_{\bf
   k}|^f_{\rm scat}.
  \label{eq26}
\end{eqnarray}
For the steady-state situation with a small current due to a static electric
field, we assume that the distribution of 
the itinerant electrons is not far away from the Fermi distribution.
The scattering effect of the spin-conserving process is introduced by the
relaxation time approximation with the average
momentum relaxation time $\tau$.
Therefore, one can linearly expand the density
matrices by considering the drift effect,
${\rho}_{\bf k}=\rho_{\bf k}^0I+{\bf
  S}_{\bf k}\cdot{\bgreek\sigma}={\rho}_k^i+e\tau
{\bf E}\cdot\nabla_{\bf k}{\rho}_{\bf k}^i$, where the isotropic density
matrices ${\rho}_k^i=\rho_{k}^{i,0}{\bf I}+{\bf
  S}_{k}^i\cdot{\bgreek\sigma}$ representing the spin polarized Fermi
distribution in the absence of the electric field. By substituting these density matrices into
Eq.\,(\ref{eq26}), the driving term and the spin-conserving 
scattering term cancel out. Further, one introduces the average spin 
relaxation time $\tau_s$ and rewrite the spin-flip scattering
as $-\tfrac{({\bf S}_{\bf k}-{\bf S}_{\bf k}^e)\cdot
  {\bgreek\sigma}}{\tau_s}$. Here, ${\bf S}_{\bf k}^e$ describes the
equilibrium spin polarization due to the spin-splitted band structure
in the ferromagnetic system.

Under the above procedures, one finally obtains the equations of the
steady-state spin polarization 
\begin{equation}
  -(2{\bf M}+\tfrac{2k_i}{m}{\bf A}_i)\times
  {\bf S}_{\bf k}-\tfrac{k_i}{m}\nabla_{\bf k}\rho_{\bf
    k}^0\cdot\nabla_{\bf R}{\bf A}_i
  =-\tfrac{{\bf S}_{\bf k}-{\bf S}_{\bf k}^e}{\tau_s},
  \label{eq27}
\end{equation}
by using the relation $\{\nabla_{\bf k}
\rho_{\bf k},\nabla_{\bf R}A_i\}=2\nabla_{\bf k}\rho_{\bf
  k}^0\cdot\nabla_{\bf R}A_i^\beta\sigma_\beta
+2\nabla_{\bf k}S_{\bf k}^\alpha\cdot\nabla_{\bf R}A_i^\alpha$. Here, we
denotes ${\bf A}_i=(A_i^x,A_i^y,A_i^z)$.

Then, the equation of the total spin polarization ${\bf S}$ can be obtained
by summing Eq.\,(\ref{eq27}) over the ${\bf k}$-space
\begin{eqnarray}
  \nonumber
  &&\hspace{-0.5cm}-2{\bf M}\times{\bf S}-\tfrac{2}{m}{\bf
    A}_i\times(\sum\nolimits_{\bf k}k_i{\bf S}_{\bf k})\\
  &&=-\tfrac{{\bf
      S}-{\bf S}^e}{\tau_s}+\tfrac{1}{m}\big(\sum\nolimits_{\bf
    k}k_i\nabla_{\bf k}\rho^0_{\bf k}\big)\cdot\nabla_{\bf R}{\bf A}_i.
  \label{eq28}
\end{eqnarray}
For the lowest order approximation, one substitutes ${\bf S}_{\bf
  k}={\bf S}_k^i+e\tau{\bf E}\cdot\nabla_{\bf k}{\bf 
    S}_{\bf k}^i$ into Eq.\,(\ref{eq28}). By considering $\sum_{\bf k}
k_i{\bf S}^i_{\bf k}=0$, one obtains
\begin{equation}
   {\bf S}-2\tau_s{\bf M}\times {\bf S}+2\tau_s{\bf
     A}_i\times(\tfrac{1}{m}e\tau E_i{\bf S})={\bf
    S}^e-\tau_s\tfrac{n}{2m}\partial_i{\bf A}_i,
   \label{eq29}
\end{equation}
in which the relation $ \sum\nolimits_{\bf k}k_i\partial_{k_j}g({\bf
  k})=-\delta_{ij}\sum_{\bf k}g({\bf k})$ is used. 
The quantity $n=2\sum_{\bf k}\rho_{\bf k}^0$ stands for the density of
the itinerant electrons. The spin polarization can then be found 
\begin{equation}
  {\bf S}=\frac{{\bf y}+{\cal A}\times {\bf y}+({\cal A}\cdot {\bf
      y}){\cal A}}{1+|{\cal A}|^2},
  \label{eq30}
\end{equation}
where ${\cal A}=2\tau_s {\bf M}-2\tau_s\tau \tfrac{e}{m}E_i{\bf A}_i$ and ${\bf y}={\bf
  S}^e-\tau_s\tfrac{n}{2m}\partial_i{\bf A}_i$. 
The equilibrium spin polarization ${\bf S}^e=\tfrac{1}{2}Pn\hat {\bf z}$ with
the value of the spin polarizability along the magnetization
direction $P$ is negative,
because ${\bf S}^e$ is anti-parallel to ${\bf M}_d$ and ${\bf
  M}$. With the current defined as ${\bf j}=e^2n{\bf E}/m$, the
transverse spin polarization is given by
\begin{eqnarray}
  \nonumber
  {\bf S}^\perp &\approx&(1+4\tau_s^2M^2)^{-1}\big(
  -\tfrac{n\tau_s}{2m}\partial_i{\bf A}^\perp_i
  -\tfrac{Mn\tau_s^2}{m}\hat {\bf z} \times\partial_i{\bf
    A}^\perp_i \\
  &&\hspace{-0.1cm}\mbox{}
  +\tfrac{\tau_s P}{e} j_i\hat {\bf z}\times {\bf A}_i^\perp
  -\tfrac{2M\tau_s^2 P}{e}j_i {\bf A}_i^\perp).
  \label{eq31}
\end{eqnarray}
Here, we have neglected the gauge field in the denominator.
We should point out that the subscript $i$ refers to the lattice
coordinate axis. In contrast, the gauge field is defined in the
rotation frame, hence one needs to transform the relevant terms back
to the lattice coordinate system.
The rotation transformations are given by $R{\bf
A}_\mu^\perp=-\tfrac{1}{2}{\bf n}\times {\partial_\mu {\bf n}}$ and
$R({\hat {\bf z}\times \bf A}_\mu^\perp)=\tfrac{1}{2}{\partial_\mu {\bf
    n}}$.\cite{kohno2} One then obtains the transverse spin polarization in the
lattice coordinate
\begin{eqnarray}
  \nonumber
  {\bf S}_l^\perp&\approx& (1+4\tau_s^2M^2)^{-1}\big[
  \tfrac{n\tau_s}{4m}{\bf n}\times{\nabla^2\bf n}
  -\tfrac{Mn\tau_s^2}{2m}\nabla^2{\bf n}\\
  &&\hspace{-0.1cm}\mbox{} +\tfrac{\tau_sP}{2e}({\bf j}\cdot\nabla){\bf n}
  +\tfrac{M\tau_s^2P}{e}{\bf n}\times({\bf j}\cdot\nabla){\bf n}
\big].
  \label{eq32}
\end{eqnarray}

Since the $s$-$d$ exchange interaction can be equivalently written as
$H_{sd}={\bf M}\cdot \langle{\bgreek\sigma}\rangle=\tfrac{2M}{M_d}{\bf
  M}_d\cdot{\bf S}$, the spin procession field induced by the $s$-$d$ exchange
interaction $-2M{\bf S}_l/M_d$ is
\begin{eqnarray}
  \nonumber
  {\bf H}^\perp&=&\tfrac{1}{M_d}(1+\beta^2)^{-1}\big[
  \tfrac{n}{4m}(1-\beta{\bf n}\times)\nabla^2{\bf n}\\
  &&\hspace{-0.1cm}\mbox{}
  -\tfrac{P}{2e}(\beta+{\bf n}\times)({\bf j}\cdot\nabla){\bf n}
  \big],
  \label{eq33}
\end{eqnarray}
with $\beta=1/(2M\tau_s)$.

We finally obtain the LLG equation with Gilbert damping
torque as Eq.\,(\ref{eq2}),
where ${\bf v}_s={\bf j} P/{[2eS_d(1+\beta^2)]}$
and 
\begin{equation}
A_{\rm ss}=n/[4m({1+\beta^2})].
\label{eq34}
\end{equation}
Here, the effective magnetic field ${\bf H}_{\rm eff}$ includes 
the isotropic and demagnetization sources\cite{konig2} as well as the
external magnetic field.

\section{Discussion}
We discuss our results based on Eq.\,(\ref{eq2}).
One notices that the third term on the right-hand side of the equation
is the current-induced torque obtained in the previous
works.\cite{szhang,kohno1,kohno2} The fourth term is independent of
the current but associated to the second-order gradient of the
magnetization. This term contains two contributions.
The one in the form $\tfrac{1}{M_d}A_{\rm ss}\nabla^2{\bf n}$
is identified as the effective spin stiffness.\cite{konig} 
In the limit
$\beta\ll1$ or $M\tau_s\gg1$, from Eq.~(\ref{eq34})
one has $A_{\rm ss}=n/(4m)$, which is consistent with the 
previous result.\cite{konig} However,  $A_{\rm ss}$
  should be modified for finite $\beta$. This stiffness is widely used in the
study on the domain wall\cite{szhang,tserk2} and demonstrated to be critical to determine
the width of the domain wall.\cite{schryer} One finds that the spin stiffness increases with
increasing the exchange coupling strength $M$, which agrees with
the previous computation.\cite{konig2}
By taking $n\sim 10^{20}$~cm$^{-3}$, $\beta\sim
1$ (Ref.\,\onlinecite{hals}), and $m=0.5m_e$ with $m_e$ representing the free
electron mass, one estimates the spin stiffness $A_{\rm ss}\sim
1$~pJ/m in GaMnAs.\cite{konig2} However, the other effective field in
the form $-\tfrac{\beta}{M_d}{\bf n}\times \nabla^2{\bf n}$ has not been studied yet. It is obvious
that this torque prevents the magnetization from varying in a
plane and induces transverse component instead. One notices that this
term has no contribution to the free energy, because it is always
perpendicular to the magnetization. Therefore, it can not be derived
from the variation of the free energy with respect to the magnetization,
which can explain the reason for missing this term in previous works.
In the following, we focus on the steady domain-wall solution of the
LLG equation in the absence of the current to illustrate the effect of
this new effective magnetic field due to the vertical spin stiffness.

In the ferromagnetic thin film or nanowire structures, one takes the
total effective magnetic field, ${\bf H}_{\rm
  eff}^{\rm tot}=Kn_x\hat {\bf x}-K_\perp n_z\hat {\bf z}+{A}_{\rm 
    eff}\nabla^2{\bf n}-\beta^\prime{A}_{\rm eff}{\bf
  n}\times\nabla^2{\bf n}$,\cite{tserk2} with 
$K$ and $K_\perp$ representing the anisotropy constant and
demagnetization field, respectively. Here, we have added stiffness
constant $A_{\rm ss}^0$ arising from the non-itinerant-electron origin (such as
dipole-dipole interaction between the localized $d$ electrons) to
describe general systems, $A_{\rm
  eff}=(A_{\rm ss}+A_{\rm ss}^0)/M_d$ and $\beta^\prime=\beta A_{\rm
  ss}/(A_{\rm  eff}M_d)$. The magnetization direction is
given by ${\bf
  n}=(\cos\theta,\sin\theta\cos\varphi,\sin\theta\sin\varphi)$. Obviously,
the ground state is the homogeneous configuration with the
magnetization pointing to the easy axis, i.e., the $\pm \hat
{\bf x}$-axis. However, the inhomogeneous configurations can also stably
exist, for example the domain wall structure. To discuss the formation
of the inhomogeneous magnetization structure, we first write the
equation of motion for $\theta$ and $\varphi$ in the absence of the current,
\begin{eqnarray}
  \nonumber
  \dot \theta+\alpha {\sin}\theta \dot\varphi&=&-\gamma
  K_\perp{\sin}\theta{\sin}\varphi{\cos}\varphi+\tfrac{\gamma
    A_{\rm eff}}{\sin\theta} 
  \partial_x({\sin}^2\theta\partial_x\varphi)\\ 
  &&\mbox{}-\gamma \beta^\prime A_{\rm
    eff}[\partial_x^2\theta-\sin\theta\cos\theta(\partial_x\varphi)^2],\\ 
  \nonumber
  -\alpha\dot\theta+{\sin}\theta\dot\varphi&=&\gamma K_\perp
  {\sin}\theta\cos\theta\sin^2\varphi+\gamma{K\sin\theta\cos\theta}\\
  \nonumber
  &&\mbox{}-\gamma A_{\rm eff}[\partial_x^2\theta-\sin\theta
  \cos\theta(\partial_x\varphi)^2]\\
  &&\mbox{}-\tfrac{\gamma {\beta^\prime} A_{\rm
      eff}}{\sin\theta}\partial_x(\sin^2\theta\partial_x\varphi).
  \label{eq35-36}
\end{eqnarray}
For the steady state, one obtains
\begin{eqnarray}
  \nonumber
  \partial_x^2\theta&=&\big[
  \tfrac{K}{A_{\rm eff}}{\sin}\theta{\cos}\theta+ 
  \tfrac{K_\perp}{A_{\rm eff}}{\sin}\theta{\sin}\varphi
  ({\cos}\theta{\sin}\varphi-{\beta^\prime} {\cos}\varphi)\\
  &&\mbox{}+(1+{\beta^\prime}^2){\sin}\theta{\cos}\theta(\partial_x\varphi)^2
  \big]/{(1+{\beta^\prime}^2)},\\
  \nonumber
  \partial_x^2\varphi&=&
  \big[\tfrac{{\beta^\prime} K}{A_{\rm eff}}{\cos}\theta
  +\tfrac{K_\perp}{A_{\rm eff}}{\sin}\varphi
  ({\beta^\prime}{\cos}\theta{\sin}\varphi+{\cos}\varphi)\\
  &&\mbox{}
  -(1+{\beta^\prime}^2)2{\cot}\theta\partial_x\theta\partial_x\varphi
  \big]/{(1+{\beta^\prime}^2)}.
  \label{eq37-38}
\end{eqnarray}
%%%GT
At ${\beta^\prime}=0$, one obtains a single wall  solution
in the $x$-$y$ plane (located at $x=x_0$), $\varphi=n\pi$, 
$\ln(\tan\tfrac{\theta}{2})=\tfrac{x-x_0}{W}$ with $W=\sqrt{A_{\rm eff}/K}$.

When ${\beta^\prime}\neq0$, the magnetization can not vary in a fixed
plane since $\theta$ and $\varphi$ are coupled. 
Unfortunately, Eqs.\,(36) and (37)
cannot be solved analytically in general. However, in the
absence of the demagnetization field ($K_\perp =0$), there is a
solution where the gradient of $\varphi$ is a constant, 
i.e., $\partial_x\varphi=\lambda$. In this case, the
equations can be written as
\begin{eqnarray}
  \partial_x^2\theta&=&\tfrac{1}{{(1+{\beta^\prime}^2)}}\big[
  \tfrac{K}{A_{\rm eff}}{\sin}\theta{\cos}\theta
  +\lambda^2(1+{\beta^\prime}^2){\sin}\theta{\cos}\theta
  \big],\\
  0&=&\tfrac{1}{{(1+{\beta^\prime}^2)}}
  \big[\tfrac{{\beta^\prime} K}{A_{\rm eff}}{\cos}\theta
  -\lambda(1+{\beta^\prime}^2)2{\cot}\theta\partial_x\theta
  \big].
  \label{eq39-40}
\end{eqnarray}
Obviously, both equations give the solution in the same form
\begin{equation}
  \ln(\tan\tfrac{\theta}{2})=\tfrac{x-x_0}{W_h},
\end{equation}
which is just the domain wall solution with the corresponding width 
$W_h^a=\sqrt{(1+{\beta^\prime}^2)/[\tfrac{K}{A_{\rm eff}}+(1+{\beta^\prime}^2)\lambda^2]}$ and
$W_h^b=2|\lambda| A_{\rm eff}(1+{\beta^\prime}^2)/({\beta^\prime} K)$, respectively.
The self-consistent condition $W_h^a=W_h^b$ determines the value of
$\lambda$  as 
\begin{eqnarray}
  \lambda=\pm\sqrt{\tfrac{K}{A_{\rm eff}}}\big[
  \tfrac{\sqrt{1+{\beta^\prime}^2}-1}{2(1+{\beta^\prime}^2)}
\big]^{\tfrac{1}{2}}.
\end{eqnarray}
The domain wall thickness is enhanced by ${\beta^\prime}$ as 
\begin{eqnarray}
  W_h=\sqrt{\tfrac{A_{\rm eff}}{K}\tfrac{2(1+{\beta^\prime}^2)}{1+\sqrt{1+{\beta^\prime}^2}}}.
\end{eqnarray}

\begin{figure}
\centering
\includegraphics[width=7.5cm]{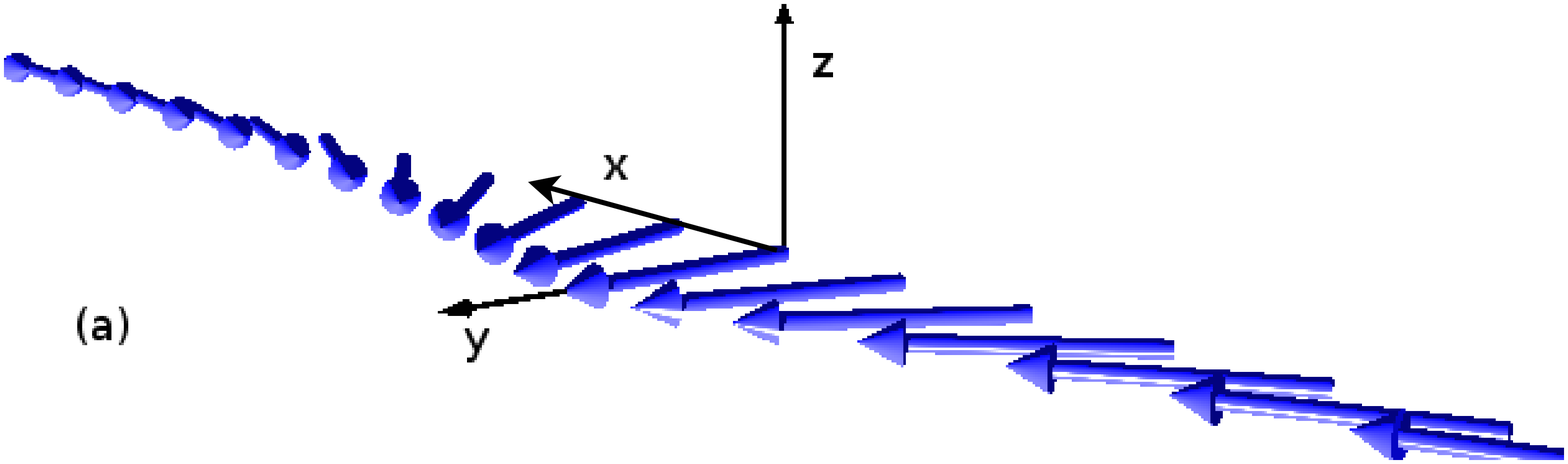}
\includegraphics[width=7.5cm]{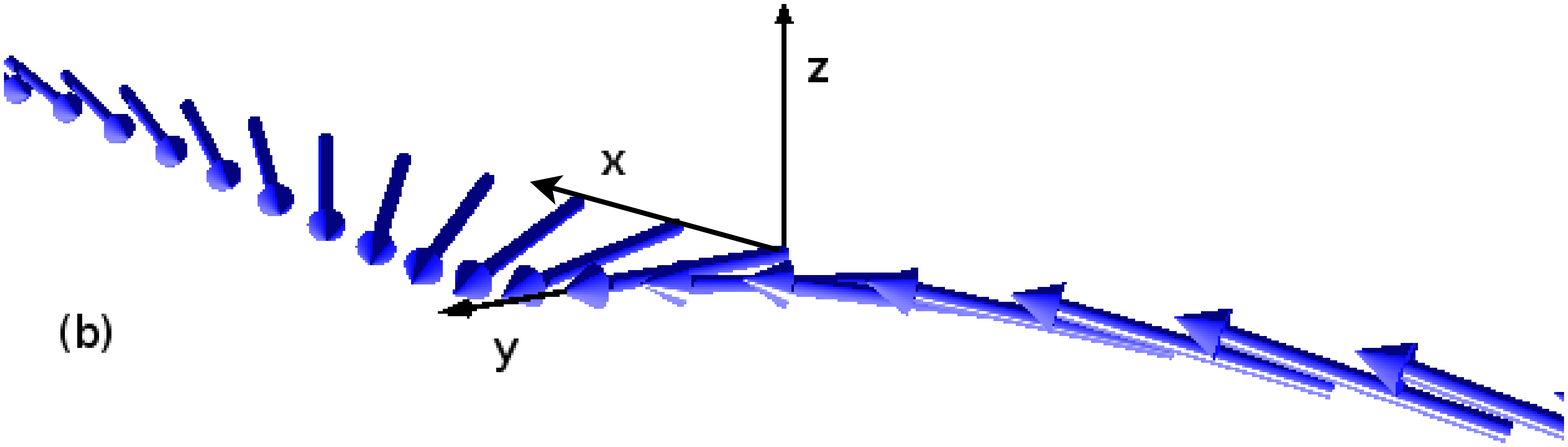}
\includegraphics[width=8.5cm,height=3cm]{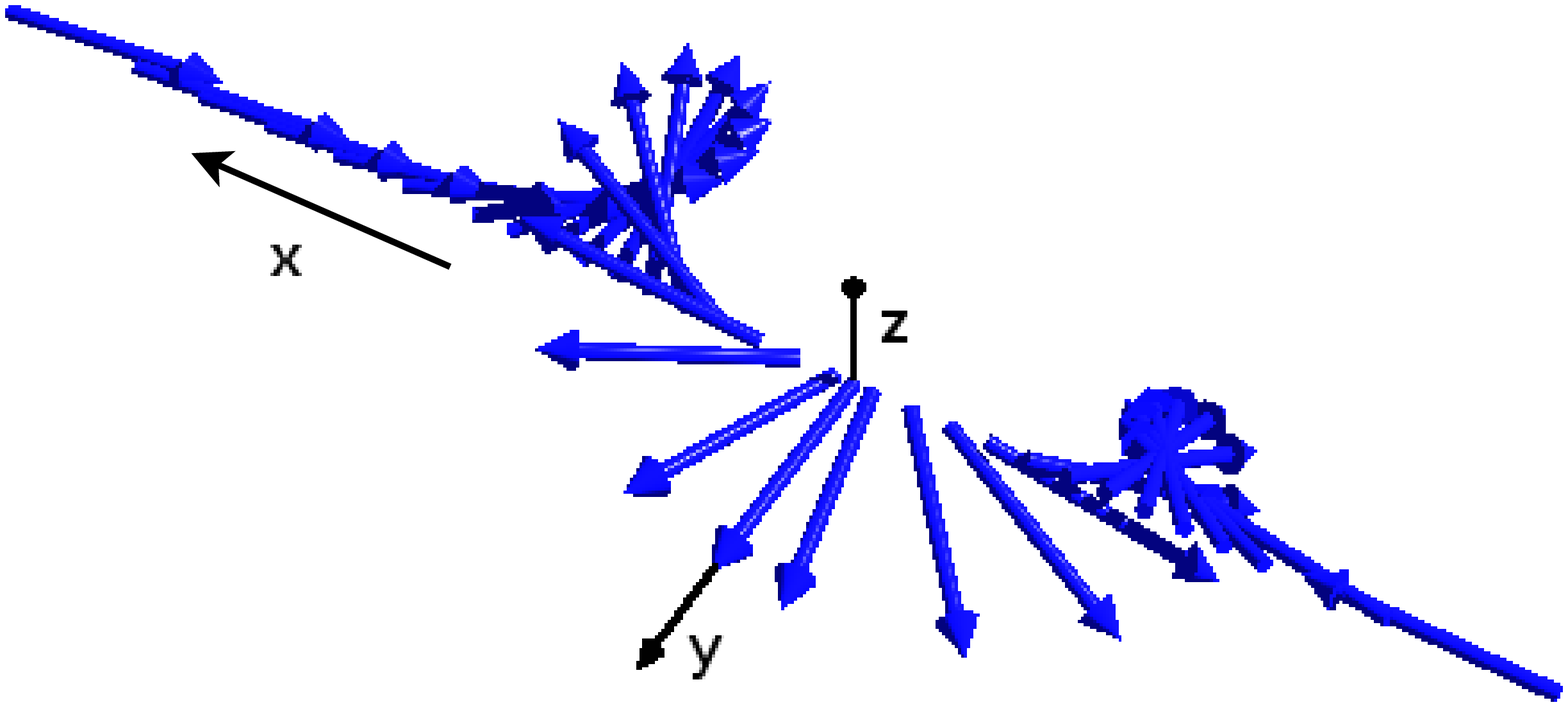}
\includegraphics[width=8.5cm]{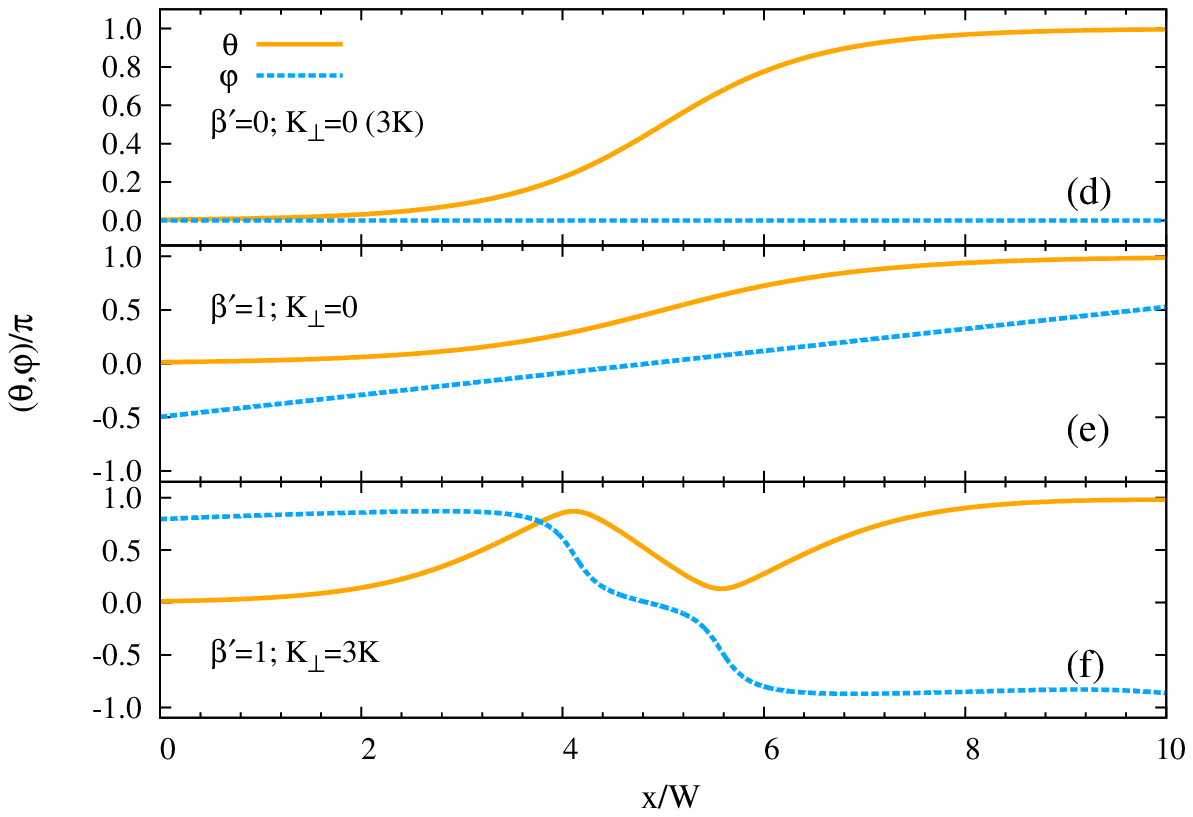}
\caption{ (Color Online) 
The variation of the direction of magnetization along 
the $x$-axis is plotted as: (a) $\beta^\prime=0$,
  $K_\perp=0$ (or $3K$); (b) $\beta^\prime=1$, $K_\perp=0$; and (c)
  $\beta^\prime=1$ and $K_\perp=3K$. By denoting the magnetization direction ${\bf
  n}=(\cos\theta,\sin\theta\cos\varphi,\sin\theta\sin\varphi)$, the
position dependences of $\theta$ and $\varphi$ are given in (d)-(f).  
$W=\sqrt{A_{\rm eff}/K}$.} 
\label{fig1}
\end{figure}

This solution indicates that the
magnetization rotates along the easy axis inside the domain wall. One
can calculate the change of $\varphi$ across the domain wall to obtain $\Delta
\varphi=\lambda W_h=(\sqrt{1+{\beta^\prime}^2}-1)/{\beta^\prime}$.
When ${\beta^\prime}\ll 1$, one finds that
$\Delta\varphi={\beta^\prime}/2 \ll1$, which indicates that $\varphi$ can be
approximated as a constant within a domain wall as shown in
 Fig.\,\ref{fig1}(a), where the magetization always lies in the
  ${x}$-${y}$ plane [see also Fig.\,\ref{fig1} (d)].
 The solution, therefore, returns to 
the N\'eel wall case.\cite{szhang}
In contrast, when ${\beta^\prime}\gtrsim1$, $\varphi$ oscillates within a wall and
the wall structure becomes more like a one-dimensional vortex with strong correlation
between $\theta$ and $\varphi$ [see Fig.\,\ref{fig1}(b) and
  (e)]. In ferromagnetic metals, $\beta$ ($\sim 0.001$-$0.01$) is
small\cite{ralph,tserk2} and $A_{ss}^0$ is large due to the strong
dipole-dipole interaction, hence the vertical spin stiffness is
unimportant. However, in ferromagnetic semiconductors, e.g., GaMnAs,
$A_{\rm ss}^0$ would be small\cite{dietl,sham} and $\beta$ can be large, $\beta^\prime
\approx \beta\sim1$,\cite{hals,jawo} therefore, the domain wall 
structure can behave like a vortex.

When $K_\perp\gtrsim1$, the behavior of $\varphi$ in a steady solution is
determined by the competition between the hard-axis anisotropy, $K_\perp
$, and the transverse stiffness is proportional to ${\beta^\prime}$. 
Then $\varphi$ tends to be locked to $\varphi= n\pi$ ($n$ is
integer) and 
deviates from being a linear function of position, forming soliton-like
structure or a staircase behavior [Fig.\,\ref{fig1}(c) and
  (f)]. In a domain wall, 
modulation  of $\theta$ occurs when $\phi$ is close to $n\pi$, but
$\theta$ tends to be close to $n'\pi $ ($n'$ is integer) when
$\varphi$  
changes in order to lower the energy cost due to the hard axis anisotropy.
Thus a domain wall has an oscillating structure as seen in Fig.\,\ref{fig1}(c) and
  (f). The number of oscillation increases as ${\beta^\prime}$ is enhanced.
As far as we find numerically, the number of oscillation also depends on the initial condition
of $\partial_x \theta$ and $\partial_x \varphi$ at the boundary of the
wall. Finally, we should point out that the domain-wall solution for
  $K_\perp\ne 0$ is the same as that for $K_\perp=0$ in the absence of
the vertical spin stiffness [see Fig.\,\ref{fig1}(a) and (d)].\cite{schryer}

As is well-known,\cite{tatara} the dynamics is
strongly affected by the structure. 
For instance,  vortex walls are easier to move than planar domain walls
in the current-driven case, because of the perpendicular component near
the vortex core.
We may therefore expect that the wall for finite ${\beta^\prime}\gtrsim1$, e.g.,
in GaMnAs,\cite{hals} would have even
lower threshold current due to the structure change arising from the
transverse exchange torque. 
Inclusion of the transverse stiffness in the micromagnetic simulations is thus crucially important in systems with strong spin-orbit interaction.

\section{Summary}
In summary, we have derived the KSBEs in
ferromagnetic systems based on the $s$-$d$ model in the presence of the
inhomogeneity of the magnetization. We analytically solved the
KSBEs and derived the spin torque due to the spin
polarization of the itinerant electrons. The current-induced spin
torque from the first-order magnetization gradient is consistent with
the previous works. We found that the
second-order gradient of the magnetization inhomogeneity gives rise
to an effective magnetic field that is perpendicular to the spin
stiffness field. This vertical spin stiffness is proportional to the
nonadiabatic parameter $\beta$. We showed that the new term modifies the domain wall
structure and causes magnetization rotation
along the easy axis. The vertical spin stiffness is expected to be
crucially important in ferromagnetic semiconductors, and needs to be
included in the LLG equation in numerical simulations on the magnetization dynamics.
\begin{acknowledgments}
This work was supported by the Natural Science Foundation of China
under Grant No.~10725417. One of the authors (G.T.) 
thanks support by a Grant-in-Aid for Scientific Research in
Priority Areas, ``Creation and control of spin current'' (Grant
No.~1948027),  the Kurata Memorial Hitachi Science and Technology
Foundation, and the Sumitomo Foundation.
\end{acknowledgments}

%\begin{appendix}
\appendix
\section{Fourier transformation of Eq.\,(\ref{eq7})}
The Fourier transformations of the functions in
Eq.\,(\ref{eq7}) respect to the relative coordinate are calculated
as follows. For simplification, we omit all the temporal coordinates
without leading to any ambiguity. 

The Green function at momentum ${\bf k}$ is defined as
\begin{equation}
  G({\bf R},{\bf k})=\int d{\bf r} e^{-i{\bf k}\cdot {\bf r}}G({\bf R},{\bf r}).
\end{equation}
Then, the electric potential energy term can be written as
\begin{eqnarray}
  \nonumber
  &&\hspace{-0.5cm}\int d{\bf r} e^{-i{\bf k}\cdot {\bf r}}U_e(1)G({\bf R},{\bf r})\\
  \nonumber
  &&=\int d{\bf r} e^{-i{\bf k}\cdot {\bf r}}
  e^{\frac{\bf r}{2}\partial_{\bf R}^U}U_e({\bf R})G({\bf R},{\bf r})\\
  \nonumber
  &&=e^{i\frac{1}{2}\partial_{\bf k}\partial_{\bf R}^{U_e}}\int d{\bf k} e^{-i\bf k\cdot
    r}U_e({\bf R})G({\bf R},{\bf r})\\
  &&=e^{i\frac{1}{2}\partial^G_{\bf k}\partial_{\bf R}^{U_e}}U_e({\bf
    R})G({\bf R},{\bf k}).
\end{eqnarray}
Terms with Hamiltonian read
\begin{eqnarray}
  \nonumber
  &&\hspace{-0.5cm}\int d{\bf r} e^{-i{\bf k}\cdot {\bf r}}H_0({\bf p}_1,{\bf r}_1)G({\bf R},{\bf r})\\
  \nonumber
  &&=\int d{\bf r} e^{-i{\bf k}\cdot
    {\bf r}} e^{\frac{\bf r}{2}\partial_{\bf R}^{H_0}}H_0(\tfrac{1}{2}{\bf
    P}_{\bf R}+{\bf p},{\bf R})G({\bf R},{\bf r})\\
  \nonumber
  &&=e^{i\frac{1}{2}\partial_{\bf k}\partial_{\bf R}^{H_0}}
  \int d{\bf r} e^{-i\bf k\cdot r}H_0(\tfrac{1}{2}{\bf P}_{\bf R}+{\bf
    p},{\bf R})G({\bf R},{\bf r})\\
  \nonumber
  &&=e^{i\frac{1}{2}\partial_{\bf k}\partial_{\bf R}^{H_0}}
  H_0(\tfrac{1}{2}{\bf P}_{\bf R}+{\bf k},{\bf R})G({\bf R},{\bf k})\\
  &&=e^{i\frac{1}{2}(\partial_{\bf k}\partial_{\bf
      R}^{H_0}-\partial_{\bf R}^G\partial_{\bf k}^{H_0})}H_0({\bf
    k},{\bf R})G({\bf R},{\bf k}),
\end{eqnarray}
and 
\begin{eqnarray}
  \nonumber
  &&\hspace{-0.5cm}\int d{\bf r} e^{-i{\bf k}\cdot {\bf r}}[G({\bf R},{\bf
    r})H_0(-\stackrel{\leftarrow}{\bf p}_2,{\bf r}_2)]\\ 
  \nonumber
  &&=\int d{\bf r} e^{-i{\bf k}\cdot
    {\bf r}} e^{-\frac{\bf r}{2}\partial_{\bf R}^{H_0}}[G({\bf R},{\bf
     r})H_0(-\tfrac{1}{2}\stackrel{\leftarrow}{\bf 
     P}_{\bf R}+\stackrel{\leftarrow}{\bf p},{\bf R})]\\
  \nonumber
  &&=e^{-i\frac{1}{2}\partial_{\bf k}\partial_{\bf R}^{H_0}}
  \int d{\bf r} e^{-i\bf k\cdot r}[G({\bf R},{\bf
    r})H_0(-\tfrac{1}{2}\stackrel{\leftarrow}{\bf P}_{\bf R}+\stackrel{\leftarrow}{\bf 
    p},{\bf R})]\\
  \nonumber
  &&=e^{-i\frac{1}{2}\partial_{\bf k}\partial_{\bf R}^{H_0}}
  G({\bf R},{\bf k})H_0(-\tfrac{1}{2}\stackrel{\leftarrow}{\bf P}_{\bf R}+{\bf k},{\bf R})\\
  &&=e^{-i\frac{1}{2}(\partial_{\bf k}\partial_{\bf
      R}^{H_0}-\partial_{\bf R}^G\partial_{\bf k}^{H_0})}G({\bf R},{\bf k})H_0({\bf
    k},{\bf R}).
\end{eqnarray}
For the integral terms, one has
\begin{eqnarray}
  \nonumber
  &&\hspace{-0.5cm}\int d{\bf r} e^{-i{\bf k}\cdot {\bf r}}\int d{\bf r}_3
  \Sigma(1,3)G(3,2)\\
  \nonumber
  &&=\int d{\bf r} e^{-i{\bf k}\cdot {\bf r}}\int d{\bf r}_3
  e^{\frac{{\bf r}_3-{\bf r}_2}{2}\partial^\Sigma_{\bf R}}\Sigma({\bf
    R},{\bf r}_1-{\bf r}_3)\\
  \nonumber
  &&\hspace{0.2cm}\times e^{\frac{{\bf r}_3-{\bf r}_1}{2}\partial^G_{\bf R}}G({\bf R},{\bf
    r}_3-{\bf r}_2)\\
  \nonumber
  &&=\int d{\bf r} e^{-i{\bf k}\cdot {\bf r}}\int d{\bf r}_3
  e^{\frac{{\bf r}_3-{\bf r}_2}{2}\partial^\Sigma_{\bf 
      R}}\int \tfrac{d{\bf k}}{(2\pi)^3}e^{i{\bf k}^\prime\cdot 
    {({\bf r}_1-{\bf r}_3)}} \Sigma({\bf R},{\bf k}^\prime)\\
  \nonumber
  &&\hspace{0.2cm}\times e^{\frac{{\bf r}_3-{\bf r}_1}{2}\partial^G_{\bf R}}\int
  \tfrac{d{\bf k}^{\prime\prime}}{(2\pi)^3}e^{i{\bf k}^{\prime\prime}\cdot 
    {({\bf r}_3-{\bf r}_2)}} G({\bf R},{\bf  k}^{\prime\prime})\\
  \nonumber
  &&=\int d{\bf r}_3 d{\bf r}\tfrac{d{\bf k}^\prime}{(2\pi)^3} 
  \tfrac{d{\bf k}^{\prime\prime}}{(2\pi)^3}e^{-i{\bf k}\cdot {\bf r}}
  \big[e^{-\tfrac{i}{2}\partial_{{\bf k}^{\prime\prime}}\partial^\Sigma_{\bf R}}
  e^{\tfrac{i}{2}\partial_{{\bf k}^\prime}\partial^G_{\bf R}} e^{i{\bf
      k}^\prime\cdot   {({\bf r}_1-{\bf r}_3)}} \\
  \nonumber
  &&\hspace{0.2cm}\times e^{i{\bf k}^{\prime\prime}\cdot
    {({\bf r}_3-{\bf r}_2)}}\big]  G({\bf R},{\bf
    k}^\prime)\Sigma({\bf R},{\bf  k}^{\prime\prime})\\
  &&= e^{\tfrac{i}{2}(\partial^G_{\bf
      k}\partial^\Sigma_{\bf R}-\partial_{{\bf
        k}}^\Sigma\partial^G_{\bf R})}
  \Sigma({\bf R},{\bf k})G({\bf R},{\bf  k}).
\end{eqnarray}
Here, the time integral $\int dt_3$ is omitted for
simplification. Similarly, one can show
\begin{eqnarray}
  \nonumber
  &&\hspace{-0.5cm}\int d{\bf r} e^{-i{\bf k}\cdot {\bf r}}\int d{\bf r}_3
  \Sigma(1,3)G(3,2)\\
  &&= e^{\tfrac{i}{2}(\partial^\Sigma_{\bf
      k}\partial^G_{\bf R}-\partial_{{\bf
        k}}^G\partial^\Sigma_{\bf R})}
  G({\bf R},{\bf k})\Sigma({\bf R},{\bf  k}).
\end{eqnarray}
In these equations, we use the notation $\partial_{\bf R}\partial_{\bf k}=\nabla_{\bf
  R}\cdot\nabla_{\bf k}$.
%\end{appendix}

\end{document}